\documentclass[acmsmall,anonymous=false,authorversion=true]{acmart}
\pdfoutput=1

\usepackage[utf8]{inputenc}
\usepackage{graphicx}
\usepackage{fixme}
\usepackage[update,prepend]{epstopdf}%
\usepackage[flushleft]{threeparttable}
\usepackage{multirow}
\usepackage{multicol}
\usepackage{tabularx}
\usepackage{enumitem}
\usepackage{color}
\usepackage{balance}
\usepackage{fp}
\usepackage{makecell}
\usepackage{booktabs}
\usepackage{pgfplots} 
\usepackage{pgfplotstable} 
\usepackage{amsmath}

\usepackage{float}
\usepackage{placeins}

\pgfplotsset{compat=1.14}

\copyrightyear{2019} 
\acmYear{2019} 
\setcopyright{acmcopyright}
\acmConference[WPES'19]{18th Workshop on Privacy in the Electronic Society}{November 11, 2019}{London, United Kingdom}
\acmBooktitle{18th Workshop on Privacy in the Electronic Society (WPES'19), November 11, 2019, London, United Kingdom}
\acmPrice{15.00}
\acmDOI{10.1145/3338498.3358650}
\acmISBN{978-1-4503-6830-8/19/11}

\begin{document}

\def\UrlFont{\smaller} 
\author{Nikolay Matyunin}
\email{matyunin@seceng.informatik.tu-darmstadt.de}
\affiliation{TU Darmstadt, CYSEC, Germany}
\authornote{\{matyunin,arul\}@seceng.informatik.tu-darmstadt.de}

\author{Yujue Wang}
\email{yujue.wang@stud.tu-darmstadt.de}
\affiliation{TU Darmstadt, Germany}
\authornote{\{yujue.wang, kristiandietmar.kullmann\}@stud.tu-darmstadt.de}

\author{Tolga Arul}
\email{arul@seceng.informatik.tu-darmstadt.de}
\affiliation{TU Darmstadt, CYSEC, Germany}
\authornotemark[1]

\author{Kristian Kullmann}
\email{kristiandietmar.kullmann@stud.tu-darmstadt.de}
\affiliation{TU Darmstadt, Germany}
\authornotemark[2]

\author{Jakub Szefer}
\email{jakub.szefer@yale.edu}
\affiliation{Yale University, USA}
\authornote{jakub.szefer@yale.edu, stefan.katzenbeisser@uni-passau.de}

\author{Stefan Katzenbeisser}
\email{stefan.katzenbeisser@uni-passau.de}
\affiliation{Chair of Computer Engineering, University of Passau, Germany}
\authornotemark[3]

\settopmatter{printacmref=true,printfolios=true} 
\pagestyle{plain} 

\definecolor{updcolor}{rgb}{0.25,0.5,0.75}
\newcommand{\updated}[1]{{\color{updcolor}{#1}}}
\definecolor{commentcolor}{rgb}{0.75,0.0,0.0}
\newcommand{\ct}[1]{{\color{commentcolor}{#1}}}

\title[MagneticSpy]{MagneticSpy: Exploiting Magnetometer in Mobile Devices for Website and Application Fingerprinting}
%


  \newcommand\varTotalDevices{80 }
  \newcommand\varAffectedDevices{56 }
  \newcommand\varNotAffectedDevices{24 }
  \newcommand\varNumOfWebsites{50 }
  \newcommand\varNumOfApps{65 }
  \newcommand\varNumOfSamples{175 }
  \newcommand\varNumOfFeatures{50 }

  \newcommand\varWebAccuracy{91\% }
  \newcommand\varAppAccuracy{90\% }
  \newcommand\varWebRecordingAccuracy{86.7\% }
  \newcommand\varTorAccuracy{37.7\%}
  \newcommand\varAgedAccuracy{2.7\% }

  \newcommand\varTracesAffectedByMovements{21\% }

  \newcommand\varOpenWorldMonitored{40}
  \newcommand\varOpenWorldNonMonitored{7,500}
  \newcommand\varOpenWorldNonMonitoredTraining{45}
  \newcommand\varOpenWorldTotal{7,700}
  \newcommand\varOpenWorldTP{68.6\%}
  \newcommand\varOpenWorldPR{92.2\%}
  \newcommand\varOpenWorldPRMonitored{90.9\%}

  \newcommand\varCURecordings{50}
  \newcommand\varCUDuration{100s}
  \newcommand\varCUInterval{1s}
  \newcommand\varCUPrecision{24,5\%}
  \newcommand\varCURecall{72.9\%}
  \newcommand\varCUAccuracy{81\%}
  \newcommand\varCUAccuracyCompared{90\% }
  \newcommand\varCUPatterns{175}

\begin{abstract}
  {Recent studies have shown that aggregate CPU usage and power consumption
  traces on smartphones can leak information about applications running on the system or
  websites visited.
  In response, access to such data has been blocked for mobile applications
  starting from Android 8.
  In this work, we explore a new source of side-channel leakage for this class of
  attacks. Our method is based on the fact that electromagnetic activity caused by mobile processors 
  leads to noticeable disturbances in magnetic sensor measurements on mobile devices,
  with the amplitude being proportional to the CPU workload.
  Therefore, recorded sensor data can be analyzed to reveal information about
  ongoing activities.
  The attack works on a number of devices: 
  we evaluated \varTotalDevices models of modern smartphones and tablets and observed the reaction of
  the magnetometer to the CPU activity on \varAffectedDevices of them.
  On selected devices
  we were able to successfully
  identify which application has been opened (with up to \varAppAccuracy accuracy) or which web page
  has been loaded (up to \varWebAccuracy accuracy).
  The presented side channel poses a significant risk to end users' privacy, as the sensor
  data can be recorded from native apps or even from web pages without user permissions. Finally, we discuss possible countermeasures
  to prevent the presented information leakage.}

\end{abstract}

\maketitle
  \renewcommand{\shortauthors}{Matyunin et al.}


\section{Introduction}
\label{sec:introduction}
Mobile devices have become ubiquitous in people's daily
activities.
According to recent studies, adults spend more than 2.5 hours per day on their smartphones or
tablets~\cite{link-smartphoneusage-1}, 
the average user runs over 30 mobile applications per month~\cite{link-smartphoneappusage-1},
while
mobile Internet traffic already exceeded desktop usage~\cite{link-mobiletraffic-2}.
Such extensive mobile usage results in an increasing amount of personal information that is
stored and processed
on mobile devices,
which increases risks of its
unauthorized or malicious misuse.
Fortunately, mobile operating system developers put a great deal of effort to limit such risks, by isolating running
applications into sandboxed environments and by introducing
permission-based access restrictions for sensitive
components~\cite{link-android-sandboxing,link-ios-security-guide}.

Nevertheless, several previous studies have shown that an attacker can exploit side-channel leakage to infer
information about applications and websites opened on a victim's mobile
device.
These leakage sources include
network traffic statistics~\cite{Zhou:2013aa,Spreitzer:2016aa,Zhang:2018aa}, power consumption traces~\cite{Chen:2017ab,Yan:2015aa}, 
CPU utilization~\cite{Zhang:2009aa,Simon:2016ab},
memory usage statistics~\cite{Jana:2012aa,Chen:2014ab,Zhang:2018aa}, and other 
information available through the~\textit{procfs} pseudo filesystem~\cite{Spreitzer:2018aa} or system APIs~\cite{Spreitzer:2018ab,Zhang:2018aa}. 
The information obtained from application and website fingerprinting can potentially reveal sensitive information
about the user, e.g., hobbies, political interests, religious beliefs, or health conditions. The more actively a victim uses the device, the more precise is the resulting user profile.

To prevent such attack vectors, operating system developers have gradually restricted access
to system resources which can reveal sensitive information.
In particular, starting from Android 7, applications cannot access pseudofiles revealing system information about other processes (e.g., \textit{/proc/[PID]})
or monitor traffic statistics of other applications~\cite{link-android-procrestriction}.
Similar access restrictions to per-process statistics are applied to applications on iOS devices starting from iOS~9~\cite{Zhang:2018aa}.
Furthermore, starting from Android 8
and on the most recent Android 9,
the
access to global system statistics available through \textit{procfs} and \textit{sysfs} is restricted~\cite{link-android-procfs-8},
preventing application and website fingerprinting attacks based on CPU utilization and power consumption traces.

In this paper, we propose an alternative source of side-channel leakage for
identifying activities running
on mobile devices, based on the reaction of magnetometer sensors to CPU activity.
It has recently been shown that peak CPU activity on a smartphone can cause a
noticeable
disturbance in magnetic
sensor measurements. In~\cite{Matyunin:2018ab},
authors utilized this observation to establish a covert channel,
by encoding a payload into binary patterns of peak and idle CPU activity and analyzing the produced sensor disturbance.
In this work, we propose to use this effect as a passive side-channel attack which aims to identify running activities.
We show that magnetometer disturbance patterns closely represent CPU workload.
Therefore, they allow to fingerprint browsing and application activity with an accuracy
comparable to the method based on observing overall CPU statistics available through \textit{procfs} before Android 8.
The proposed method does not require any user permissions at the moment.
As a result, any application installed on a device can
infer running applications or visited websites,
unnoticeable to the end user.
Furthermore,
the magnetometer can now be accessed within web pages using the recently-introduced
Generic Sensor API~\cite{link-generic-sensor-api}. In this case,
the attack does not even require an installed malicious application. Instead,
a web page under the attacker's control can establish
fingerprinting of other web pages or applications.

We have examined \varTotalDevices popular smartphones and tablets, and have found that magnetometers on \varAffectedDevices of them are affected by CPU activity.
For these devices, we created
a classifier which analyzes disturbances in recorded sensor measurements to identify
activities on a device.
In practical scenarios, we are able to
identify an opened
website with an accuracy of up to \varWebAccuracy for a set of \varNumOfWebsites popular websites.
We were also able to identify a running application with up to \varAppAccuracy accuracy for a set of \varNumOfApps candidate applications.
In all cases, the accuracy is significantly higher than the baseline accuracy obtained from random guessing,
and is comparable to the approach based on analyzing \textit{procfs} information.
Therefore, the presented
side channel can pose significant privacy risks to end users.

\subsection{Contributions}
Our contributions can be summarized as follows:
\begin{itemize}
\item We investigate the reaction of magnetic sensors
to varying CPU activity on \varTotalDevices different smartphones
and tablets in cloud and lab environments.
To the best of our knowledge, our work is the first to test this side channel 
on a large number of devices running both Android and iOS platforms.
\item We propose to exploit this side channel for application and website fingerprinting on mobile devices. We show how to extract information
from magnetometer disturbances, evaluate the classification performance under realistic conditions, and discuss possible countermeasures.
\item We show that our method provides classification accuracy comparable to techniques based on \textit{procfs} leakage, but works in presence of security enhancements implemented in
the latest mobile operating systems,
and can be run in both in-app (malicious app) and in-browser (malicious web page) scenarios.
\end{itemize}



\section{Background}
\label{sec:background}

In this section,
we provide background information, describing
the use of magnetometers in mobile devices
and show the reaction of magnetometers to electromagnetic activity caused by the CPU.

\subsection{Magnetometers} 
\label{sec:background:sensors}

Most modern smartphones and tablets are equipped with magnetic sensors, also called magnetometers.
These sensors measure the ambient geomagnetic field intensity for all three physical axes in units of micro Tesla, usually by utilizing the Hall effect~\cite{Cai:2012aa}.
Normally, they are used to estimate the orientation of the device relative to earth’s magnetic north
and in this way act as digital compasses, e.g., to show the user’s current direction in navigation applications.

In Android and iOS native applications, 3-axis magnetometer values can be retrieved using
the Sensor~\cite{link-native-android-sensor-api} and the Core Motion~\cite{link-native-ios-sensor-api} frameworks, respectively.
Depending on a device, the sampling rate is limited by
the operating system to
50--100Hz.
Access to the sensor does not require any explicit permissions, and therefore, any installed application
can read
sensor measurements without user attention.
In web applications, magnetometer data can be accessed using the
recently introduced
Generic Sensor
API \cite{link-generic-sensor-api}, currently in W3C Working Draft status.
The API is
available in Google Chrome and Opera web browsers, but at the moment access to magnetometer requires a configuration
flag to be explicitly enabled~\cite{link-generic-sensor-api-chrome}.
In comparison to native APIs,
the Generic Sensor API
has additional limitations:
First, the data can be accessed only from the
foreground tabs,
and only for web pages opened using HTTPS.
Second, the sampling rate is
limited to 10~Hz. Nevertheless, we present attack scenarios which work even in the presence of these
limitations (Section~\ref{sec:scenarios}), while our experiments (Section~\ref{sec:evaluation:sampling-rate}) show that
a sampling rate of 10~Hz does not prevent the side channel.

\begin{figure}[!t]
  \centering
  \includegraphics[trim={0mm 0mm 0mm 0mm},clip,width=\textwidth]{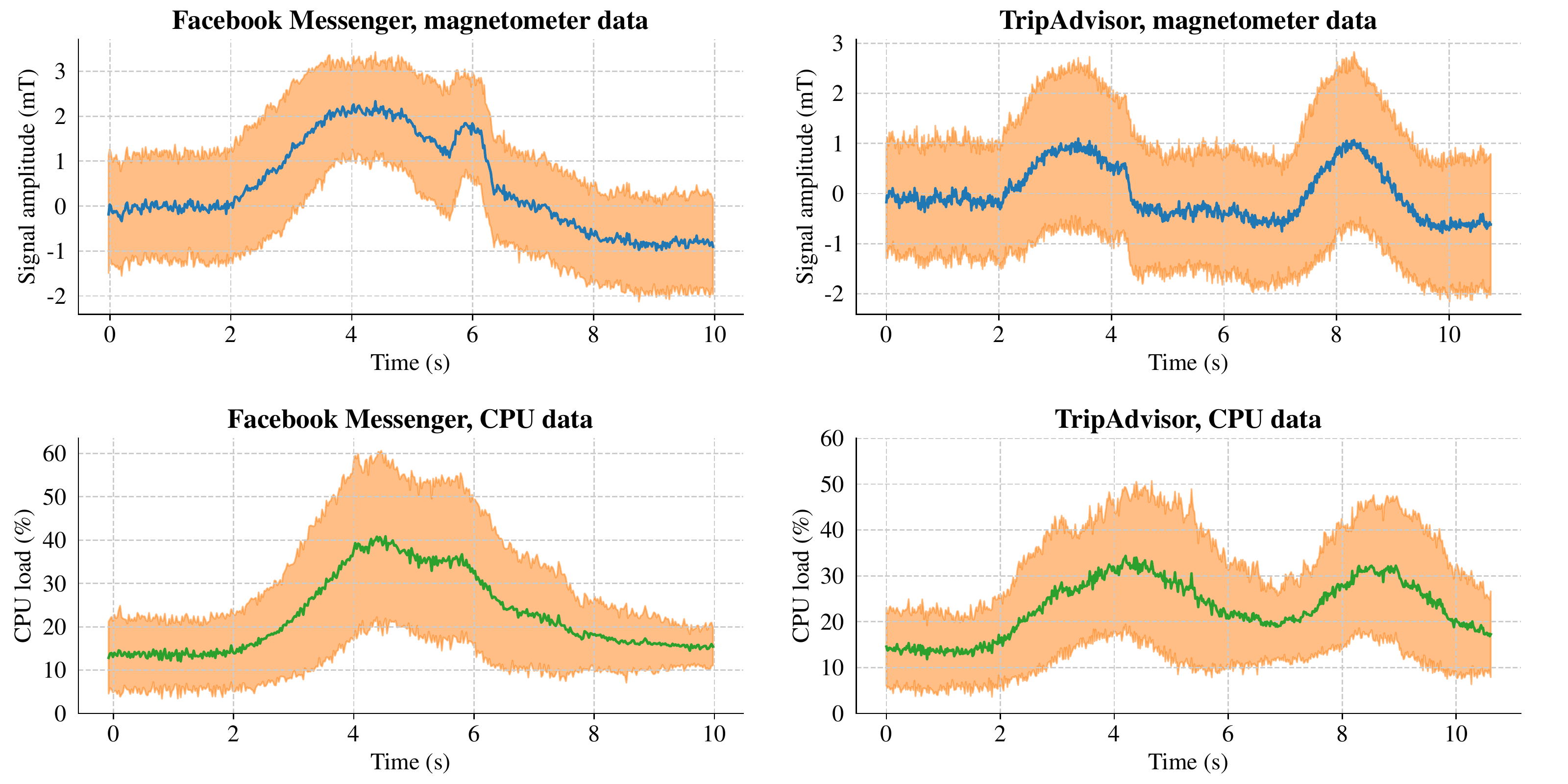}
    \caption{Examples of CPU utilization and magnetometer measurements, recorded during opening two applications
    on a Google Pixel 2 smartphone. Plots represent mean and standard deviation for \varNumOfSamples samples.
    The CPU and sensor data are visually correlated with each other for each application and are significantly different between applications.}
  \label{fig:activity-example}
\end{figure}

\subsection{Sensitivity of magnetometers to CPU activity}
As it has been discovered in prior work~\cite{Guri:2018ab,Matyunin:2016ab}, the magnetometer on mobile devices is susceptible to the
electromagnetic radiation emanated from electronic devices located nearby.
In particular, high CPU workload on a device typically requires more power, which results in a higher produced electromagnetic field.
Matyunin~et~al.~\cite{Matyunin:2018ab} showed that this effect is observable on a smartphone: Very high CPU activity on a Nexus 5X smartphone (close to 100\% of the CPU load) led to a noticeable peak in magnetometer measurements.

In this work, we further investigate the reaction of magnetometers to CPU activity on mobile devices. We have observed that on many smartphones the pattern of the sensor disturbance accurately follows the pattern of the CPU utilization.
The reasons for this are the following: On one hand, the
CPU is one of the most power-consuming components of the device~\cite{Carroll:2010aa}.
The screen and GSM module can consume more power, but their consumption remains comparably stable during normal usage.
On the other hand, mobile processors are optimized to consume minimum power under low or idle activity.

At the same time, different applications or websites 
require different amounts of CPU resources when running.
As a result, CPU utilization traces,
as well as the corresponding sensor disturbance,
can contain distinct patterns which uniquely identify the activity. Figure~\ref{fig:activity-example} shows CPU utilization traces recorded on a smartphone for two
applications,
in combination with
magnetometer readings recorded at the same time.
The 
patterns in the corresponding CPU and sensor
measurements are visually correlated for each application, they are stable within multiple recordings, but
distinct for two different applications. In this work, we show that an adversary can effectively extract information out of such recorded magnetometer measurements, and use it to perform application and website
fingerprinting on a victim's device.


\section{Attack scenario}
\label{sec:scenarios}

In this section, we discuss two considered attack scenarios, \textit{in-app} and \textit{in-browser},
discuss their limitations and elaborate on the considered assumptions.

In the \textit{in-app} scenario, the victim installs an attacker-controlled
application
on his or her device. This application
does not require
privileged access rights
from the system and does not have additional user permissions,
apart from access to the Internet, granted by default.
Therefore, malicious code can
be hidden in any application which victims are likely to install.
This application can be sandboxed according to the latest
Android and iOS security enhancements. In particular, this application does
not have any information about other running applications or network
traffic, and does not have access to system resources (e.g., over
\textit{procfs} or \textit{sysfs}). The attacker only has access 
to zero-permission sensor information.

In the \textit{in-browser} scenario, a victim opens a web page
under the attacker's control. The web page either fully belongs
to the attacker, or contains components from an
attacker-controlled server, similarly to the case when websites
include third-party code from advertisement
and analytics
services.
Such third-party components can be present on thousands of websites,
which makes this scenario comparably even more scalable.
Similarly, we assume that this web page is sandboxed by the browser from
other web pages, processes and system resources. 

In both scenarios, an attacker constantly collects 
magnetometer readings and tries to identify opened applications or
websites by applying a supervised learning approach. To achieve
this, the attacker needs to perform a training phase, which requires gathering
a sufficient set of labeled traces for each visited website or
application. A powerful attacker can perform learning on a large number
of devices which he or she owns or accesses using cloud testing platforms, such as AWS Device Farm~\cite{link-awsfarm}.
On a victim's device, the
attacker only collects traces to be classified during the testing phase and sends them to
a server. The attacker may additionally send 
information about the victim's device to a server,
to match the victim's device with same model of the device that the attacker trained on,
as model-specific classification
has a higher success rate (we evaluate this in
Section~\ref{sec:evaluation:fingerprinting}). On Android, the device model
is freely accessible by applications through the \textit{Build.MODEL}
property; on iOS, it can be retrieved using the \textit{UIKit} framework and \textit{uname} system call. In the in-browser scenario, this device model can be obtained from the User-Agent HTTP
header~\cite{link-user-agent}.

Alternatively, for the website fingerprinting case, an attacker can
perform the learning phase directly on the victim's device. For
this purpose, a malicious application can embed an
invisible WebView~\cite{link-android-webview} component to open all websites from 
the training dataset and send
the labeled sensor data to an attacker-controlled server. Although such an approach would
provide the most precise device-specific training dataset, in this
case,
the application
needs to be actively
used in the foreground by the victim for a significant amount of time.

    \begin{table*}[t]
    \small
    \centering
    \pgfplotstableread[col sep = comma]{tables/scenarios.csv}\datascenarios
    \caption{Comparison between the in-app and in-browser attack scenarios}
    \begin{threeparttable}
    \pgfplotstabletypeset[
      col sep = comma,
      every head row/.style={before row={}, after row=\midrule, },
      fixed,
      fixed zerofill={true},
      precision=2,
      multicolumn names=l,
      empty cells with={---},
      columns={property,local,remote},
      columns/property/.style={column name={}, string type, column type={p{0.30\textwidth}}},
      columns/local/.style={column name={In-app scenario}, string type, column type={p{0.33\textwidth}}},
      columns/remote/.style={column name={In-browser scenario}, string type, column type={p{0.3\textwidth}}},
      row predicate/.code={}%
    ]{\datascenarios}
    \end{threeparttable} 
    \label{table:scenarios}
    \end{table*}

\subsection{Applicability of the scenarios}
\label{sec:scenarios:applicability}
\label{sec:scenarios:background-limitations}

The in-app and in-browser attack scenarios
also differ regarding their applicability. Due to technical limitations
of the Generic Sensor API  discussed in Section~\ref{sec:background:sensors},
the magnetometer can only be accessed from foreground browser
tabs. Therefore, the in-browser scenario can only be used to 
identify either \textit{background} activities, or websites and applications
opened side by side with the recording web page, in so-called
\textit{split screen} mode.

In the in-app scenario, the time frame
during which the
malicious
application can gather magnetometer traces depends
on the platform and OS version. Starting from Android 8, the background
execution of applications is limited to several minutes after the last
user interaction with the application~\cite{link-android-background-execution}.
In the newest Android~9, sensors cannot be accessed in the background
by default~\cite{link-android-9-sensors}.
To be able to continuously record sensors in the background on Android 8 and 9, the attacker
needs to declare a so-called \textit{ForegroundService}~\cite{link-native-android-sensor-api}, which results
in a visible user notification.
This notification, 
however, can be masqueraded as a seemingly benign functionality which
needs to be constantly running, e.g., a fitness activity tracker.

Similarly to Android 8 and newer, execution of iOS applications is
suspended shortly after being moved to the background (starting from iOS~5).
Furthermore, the iOS platform does not provide a functionality 
to keep the background application active for an
arbitrary amount of time.
However, execution time can be granted to background applications when they perform specific set
of actions, e.g. playing audio, receiving location updates, processing updates
from a server or reacting to remote push
notifications~\cite{link-ios-background-execution}.

Nevertheless, in all cases, the attacker
would be able to classify applications or webpages opened shortly after the victim
leaves a malicious application, classify background activities, or activity
opened in split screen mode. Table~\ref{table:scenarios}
summarizes the differences between the in-app and in-browser scenarios.

\subsection{Additional assumptions}
\label{sec:scenarios:assumptions}
Following other works on website and application fingerprinting,
we discuss several additional assumptions~\cite{Juarez:2014aa}.
In this section, we reason about these assumptions 
with regard to our scenarios, and show that many of them can be encountered
on modern mobile platforms, in comparison to traditional desktop systems.

First, it is typically assumed that 
users open applications (websites) sequentially and
have only a single \textit{active} application (website) open at a time. This assumption
is reasonable for our scenarios, as
on mobile devices
a user can not keep more than two applications in the foreground at a time, with two only in 
split-screen mode. Furthermore,
modern mobile browsers significantly limit 
JavaScript execution in background tabs or even completely prevent it, to reduce
power consumption. 
As a result, in a general case only one application remains active at a time, and, in case
of a web browser, only one tab can be active.

Second, we assume that there is no user-invoked activity in the background.
As described in Section~\ref{sec:scenarios:background-limitations}, modern Android and iOS systems 
limit execution time of background processes. We confirmed that these limitations
result in low average background activity. We performed the test measurement of the average
CPU activity over a period of 24 hours on two devices
with numerous applications installed and the recording application in the foreground.
As a result, we obtained the average CPU utilization of only 
1.9\%, with the standard deviation of 1.7\%.
Furthermore, in the course of our experiments we did not take any measures to
specifically prevent background activity.
We performed our measurements on unmodified smartphones, with up to 60 additional
popular applications installed.
These applications could potentially generate CPU noise in the background during 
the continuous recording (over 30 hours of recording per device and tested scenario). 
Nevertheless, the high classification rates show that these activities 
do not significantly affect the recording traces.
Overall, we can expect that the impact of background acitivtiy on the classification is low.

Third, we present evaluation results under the assumption that websites
(applications) do not change over time. As observed in our experiments and other works
(e.g., see~\cite{Juarez:2014aa,Yang:2017aa}),
this assumption does not hold for websites, and the attacker needs
to periodically re-run the learning phase. However,
we observed that traces from applications remain stable unless they get updated. 
In addition, configuration options of the browser are comparably
limited on mobile devices, so it is easier for the attacker to
replicate the user client-side settings.

Finally, it is generally assumed that the attacker can detect the beginning
and end of each activity to be classified. In practice, this can be hard to
achieve: In our case, \textit{any} CPU activity performed on a device can cause 
magnetic disturbances.
As one potential solution, we show in Section~\ref{sec:methodology:continuous-usage}
that the attacker can identify potential time points when the target activity
could have started by computing the cross-correlation with the predefined pattern,
and run the classification only at these specific points.

Apart from these assumptions, typically addressed in works on website and application
fingerprinting, in this work
we additionally assume that the victim is not actively moving
the device, as movements affect magnetometer data.
In Section~\ref{sec:evaluation:movements}, we evaluate the impact of minor movements on 
the classification accuracy when the smartphone is being held in hand.
Furthermore, in Section~\ref{sec:methodology:movements}, we propose
an approach how the attacker can
identify and filter out sensor readings which are disturbed by movements.

As a result, we believe that our scenarios are realistic
under given assumptions.

\section{Attack details}
\label{sec:methodology}

In this section, we discuss implementation details about how data was collected,
its pre-processing, feature extraction and classification,
describe approaches to identify the target activity in the
continuous measurement stream and to identify traces disturbed by device movements.

\subsection{Data collection}

To collect a large set of labeled
traces in the learning phase, we trigger opening applications and
websites from our datasets in an automated
and controllable
way,
using the Android Debug Bridge (\textit{adb})~\cite{link-android-adb} tool
from the Android SDK.
Our service script opens each application from the dataset,
waits for a predefined duration, and closes the target application. Similarly,
for website fingerprinting, the service script opens the Chrome
browser and the corresponding website.
Additionally, we 
implement opening websites in a separate application with an
embedded WebView component. It allows us to evaluate the website fingerprinting in the cloud
testing platforms, such as the AWS Device Farm~\cite{link-awsfarm}.
These platforms
allow developers to test mobile applications remotely on multiple devices. However, they
do not provide access to devices through the \textit{adb}. Therefore, we could not evaluate
the application fingerprinting or use the Chrome browser on these platforms.

To collect resulting magnetometer disturbance traces, we implemented
an Android application which runs in the background, records 
3-axis magnetometer data, and sends it to the attacker-controlled server. Similarly,
for the in-browser scenario, we implemented a web page which records the
sensors using the Generic Sensor API in the mobile Chrome browser,
and sends the data to the server.
As a result, for each opened application or website, the server
receives labeled (in the learning phase) or unlabeled (in the
testing phase)  sensor measurements.

\subsection{Data preprocessing}
Subsequently, we convert the raw 3-axis data trace into a discrete-time
one-dimensional trace.  For this purpose, we
apply Principal Component Analysis~\cite{hotelling1933analysis}
to the data, choosing the first component as the result.
The resulting data represents the one-dimensional axis with the highest data
variance. Assuming that the orientation of the device is not changed significantly
and that the ambient magnetic field together with EM
noise is constant at a given point in time, this variance represents
the vector of the EM emanation caused by the CPU.
The disturbance in the
one-dimensional trace can be directed above or below the baseline level.
Therefore, 
we add both the original recorded trace and its inverse with regard to the baseline 
to the dataset in the training phase of the classifier, considering both traces as representations of the corresponding CPU pattern.

Finally, we
normalize the result to the range [0--1], so the resulting values do not depend on the
maximum possible amplitude of the disturbance (which
is device-specific, see Section~\ref{sec:evaluation:leakage}).
Instead, the result contains information about the ``shape'' of the pattern,
which represents the unique CPU activity pattern.

\subsection{Feature extraction and data classification}
Finally, we divide the resulting normalized discrete-time values of the axis with the biggest
variance into equal-size overlapping intervals (bins)
and calculate the mean value within each bin. These mean values are
used as features for classification. To classify the traces, we
use a Random Forest~\cite{Breiman:2001aa} machine learning classifier,
as it outperforms other algorithms in our experiments in terms of resulting classification
accuracy.
We split the dataset into training set (80\%) and test set (20\%).
The 5-fold cross-validation is performed on the training set to select optimal hyperparameters using the grid search,
which include
the number of estimators in the forest, the maximum number of features, maximum depth of the tree, and minimum impurity decrease.
The test set was only used to compute the accuracies when evaluating the classifier in our experiments.

In our experiments, we use the RandomForestClassifier 
from the \textit{scikit-learn} library~\cite{scikit-learn} to perform classification.
The values of the hyperparameters selected after the cross-validation are: \textit{n\_estimators} = $1100$, \textit{max\_features} = \textit{log2}, \textit{max\_depth} = $50$, \textit{min\_im\-purity\_decrease} = $0.0001$. Other hyperparameters are kept as default.

\subsection{Identifying target activity during continuous usage}
\label{sec:methodology:continuous-usage}
As we discussed in
Section~\ref{sec:scenarios:assumptions}, the attacker is assumed
to know the beginning of the activity to be classified.
In our case, the attacker
needs to continuously monitor magnetometer
disturbances, which can be caused by \textit{any} application.

However, if the practical goal of the fingerprinting is to identify
whether the victim opens a particular target application or a website
(or set of or websites),
we propose the following approach to reduce the
amount of data to be processed
by the classifier.
First, the attacker can compute an averaged CPU activity pattern for the
target 
application or website
by computing mean values along multiple traces for this activity
(known from the learning phase).
Then, this pattern can be 
used to calculate the cross-correlation with the continuously 
recorded data using the following formula:
$$c_{tp}[k] = \sum_n t[n+k] p[n],$$
where $t$ is a recorded discrete trace and $p$ is the computed pattern.
If the target activity was produced within the recorded interval,
a strong peak is present in the cross-correlation result
at the corresponding time point.
In practice, however, due to noise and slight changes in the produced
activity patterns,
cross-correlation results will not have a single strong peak, but multiple potential peaks. 
However,
due to similarity in actual and averaged patterns,
one can expect that the actual time point corresponds to one of these peaks.
Therefore, the classification can be run only at time points where peaks are present in the cross-correlation result
with a predefined threshold. This threshold sets a trade-off between the number of peaks and the accuracy of peak detection.
We evaluate this approach in Section~\ref{sec:evaluation:continuous-usage}.

Interestingly, for website fingerprinting, an attacker can also perform this step on recorded data
to first detect the web browser application to be opened (as an application fingerprinting task), and then classify the recorded interval
after the browser was opened.

\subsection{Identifying device movements}
\label{sec:methodology:movements}
If the victim rotates the device, a corresponding change in the global orientation and relative direction to the magnetic north
will cause a shift in magnetometer readings along three axes. In this case, the PCA-based trace will no longer represent disturbance exclusively caused
by CPU activity.
To identify and filter out traces which are affected by movements, we propose to analyze the rotation rate measurements
from the gyroscope sensor simultaneously with the magnetometer.
Access to gyroscope also does not require permissions, and its data is not affected by the CPU activity.
Therefore, the attacker can use gyroscope readings to estimate if the device has been significantly moved.

More specifically, we propose two criteria for identifying traces affected by movements. The first criterion is 
the mean amplitude of the rotation rate along all three axes, which indicates the overall presence of movements within the recorded interval.
The second criterion is the highest amplitude of the rotation rate, which indicates abrupt change in orientation.
If the value computed for any of two criteria exceeds a predefined threshold, the recorded trace
is considered to be affected by movements, and the trace can be ignored during the classification.
In Section~\ref{sec:evaluation:movements}, we evaluate this approach for the smartphone being held in hand. 

\section{Evaluation}
\label{sec:evaluation}

In this section, we identify devices on which magnetometers
are affected by the CPU, 
evaluate the classification performance, 
show the success rate of capturing the target activity, and
investigate the impact of minor movements.

\subsection{Information leakage}
\label{sec:evaluation:leakage}

\begin{figure}
 \centering
            \includegraphics[trim={0mm 0mm 0mm 0mm},clip,width=\columnwidth]{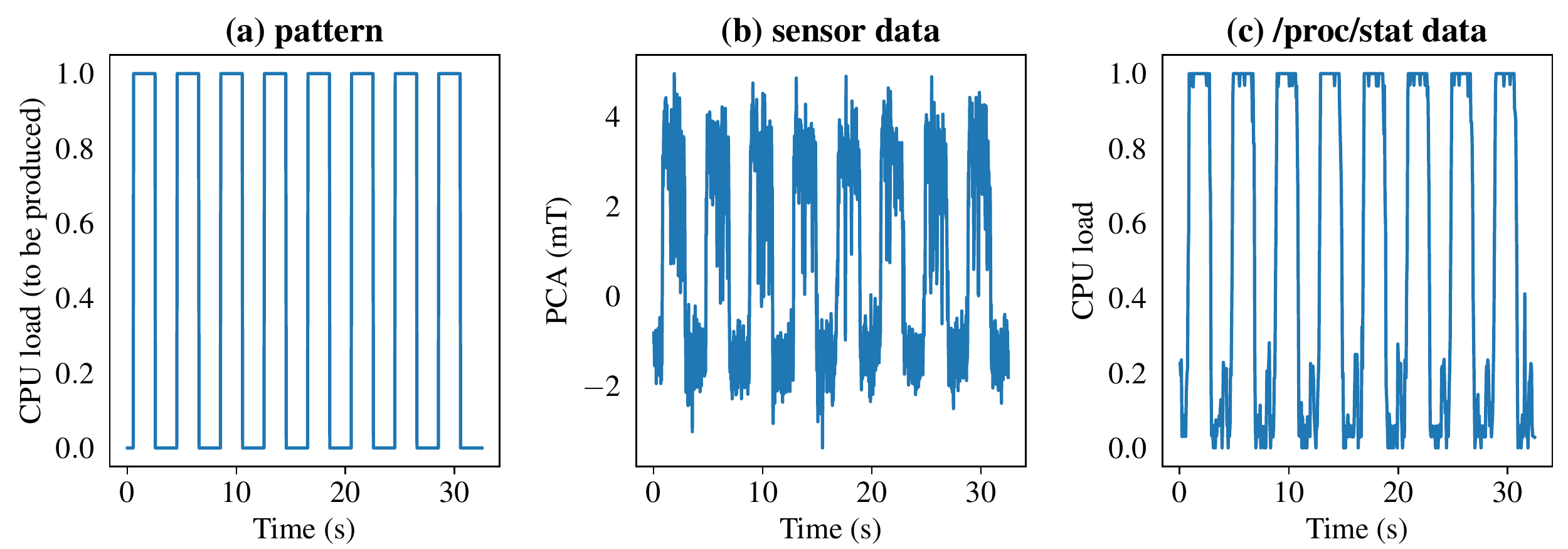}
            \captionof{figure}{Example of the
            expected CPU pattern to be produced \textbf{(a)}, recorded sensor data \textbf{(b)}, and actual CPU pattern recorded
            through \texttt{/proc/stat} \textbf{(c)}.}
            \label{fig:snr-example}
\end{figure}

In this experiment, we examined whether the magnetometer readings on mobile
devices are affected by the CPU workload. For this purpose, we produced a
predefined CPU activity pattern on a device and analyzed
resulting sensor disturbances. The pattern
consists of alternating high and low CPU loads lasting for 2 seconds.
To produce high loads, we concurrently ran so-called busy waiting loops
in a number of threads,
equal to the number of available logical cores on a device,
utilizing up to 100\% of the CPU time.
To produce low loads, we paused the execution.

Afterwards, we calculated the correlation between
this pattern and recorded
measurements. If the device
runs Android 7 or iOS, we were able to additionally
calculate the correlation coefficient with the actual produced
CPU activity pattern, recorded using \textit{/proc/stat} or \textit{host\_processor\_info}, respectively.
Some examples of predefined pattern and corresponding
magnetometer and \textit{/proc/stat} recordings are illustrated in
Figure~\ref{fig:snr-example}.
We also measured the Signal to Noise
Ratio (SNR), i.e., the ratio between the average amplitude of the disturbance caused by the high
CPU load and the standard deviation of measurements without CPU
activity. It allows us to estimate how robust the produced disturbance is
against environmental and intrinsic noise.
To evaluate a large number of devices, we conducted
measurements using two cloud platforms, Visual Studio App Center~\cite{link-appcenter} and
AWS Device Farm~\cite{link-awsfarm}. We selected all available devices running Android 7 
or higher, and all devices running iOS 11 or higher. Additionally, five devices were used in the
lab in a typical office environment.
We could not control the
environment of the devices in the cloud (such as noise), and tested them as is.

      \begin{table*}[t!]
      \caption{
      Selection of devices on which sensor measurements 
      correlate with CPU activity. The full list of affected devices is presented in Appendix~\ref{sec:appendix}.
       The table shows the cross-correlation between sensor data
      and expected CPU activity pattern (Corr.Pattern); for Android$\le$7 and iOS, also between sensor data and
      \textit{actual} CPU loads (Corr.CPU), as well as SNR ratios.
      }
      \label{table:snr-affected-part}
            \centering
            \small
      \pgfplotstableread[col sep = comma]{./tables/snr-3-affected-part.csv}{\datasnraffected}
      \begin{threeparttable}
      \pgfplotstabletypeset[
        col sep = comma,
          every head row/.style={before row={%
                                    \multicolumn{1}{c}{Smartphone}
                                 & \multicolumn{1}{c}{Setup$^\text{a}\!\!$}
                                 & \multicolumn{1}{c}{Magnetometer$^\text{c}\!\!$}
                                 & \multicolumn{2}{c}{Correlation}
                                 & \multicolumn{1}{c}{SNR,} \\
                               },
                               after row=\midrule,
        },
        fixed,
        fixed zerofill={true},
        precision=2,
        multicolumn names=l,
        empty cells with={---},
        every row no 0/.style={
          before row={\multicolumn{6}{l}{\textbf{Android}}\\},
        },
        every row no 16/.style={
          before row={\midrule\multicolumn{6}{l}{\textbf{iOS}}\\},
        },
        columns={smartphone, setup, magnetometer, correlation, correlation-cpu, snr},
        columns/smartphone/.style={column name={}, string type, column type={p{0.30\columnwidth}}},
        columns/setup/.style={column name={}, string type, column type={p{0.10\columnwidth}}},
        columns/magnetometer/.style={column name={}, string type, column type={l}},
        columns/correlation/.style={column name={~Pattern~}, column type={r}},
        columns/correlation-cpu/.style={column name={~CPU$^\text{b}\!\!$~}, column type={r}},
        columns/snr/.style={column name={dB}, precision=1,column type={r}},
        row predicate/.code={}%
      ]{\datasnraffected}
      \begin{tablenotes}
      \begin{footnotesize}
      \item[a] V --- Visual Studio App Center; A --- AWS Device Farm; L --- lab
      \item[b] CPU utilization data is available only on devices running Android $\le$ 7 (over \textit{/proc/stat})
      and iOS (over \textit{host\_processor\_info}).
      \item[c] For Android devices, information is available from the \textit{Sensor} API. For iOS devices, information from publicly available online resources is used.
      \item[] 
      \end{footnotesize}
      \end{tablenotes}
      \end{threeparttable} 
      \end{table*}

We found that magnetometers on \varAffectedDevices out
of \varTotalDevices devices are affected
by the CPU activity.
Results for selected devices are shown in
Table \ref{table:snr-affected-part}, the full list is provided in Appendix~\ref{sec:appendix}.
On these devices, the disturbance
clearly correlates with the CPU activity (with correlation scores over
80\% on average).
On most of the devices, the signal exceeds noise.
In further experiments, we confirmed that a SNR of $\approx$4dB is
sufficient to establish fingerprinting. Magnetometers on other \varNotAffectedDevices
devices,
listed in Appendix~\ref{sec:appendix},
however,
were not affected by CPU activity.

As one can see from both tables, the
sensor model does not indicate whether the magnetometer is affected
by the CPU: For example, sensors
AKM AK09915 and AKM AK0991X can be found on both affected and not
affected devices.
We believe that the reaction mostly depends on
the physical location of the sensor with regard to the CPU and
power wires, and applied shielding.

As a result, we believe that the attack is practical, since modern
popular devices (e.g., recently released smartphones Google Pixel 3, 
Samsung Galaxy S10, and iPhone XS) are all affected.

\subsection{Classification results}
\label{sec:evaluation:fingerprinting}

      \pgfplotstableread[col sep = comma]{tables/classification-basic-combined.csv}\dataclassificationbasic
      \begin{table}[t]
      \caption{Classification accuracy for website and application fingerprinting in the in-app and in-browser scenarios
      compared to classification using \texttt{/proc/stat} data. Traces have been
      collected on a Google Pixel 2 smartphone in the lab environment.}
      \label{table:classification-basic}
      \centering
      \small
      \begin{threeparttable}
      \pgfplotstabletypeset[
        col sep = comma,
        every head row/.style={before row={%
                                    \multicolumn{1}{c}{Dataset}
                                 & \multicolumn{1}{c}{Setup}
                                 & \multicolumn{1}{c}{Browser}
                                 & \multicolumn{1}{c}{Sampling}
                                 & \multicolumn{1}{c}{Accuracy,} \\
                               },
                               after row=\midrule,
        },
        every row no 0/.style={
           before row={\multicolumn{5}{c}{\textbf{Website fingerprinting}}\\}
        },
        every row no 3/.style={before row={}, after row=\midrule, },
        every row no 4/.style={
           before row={\multicolumn{5}{c}{\textbf{Application fingerprinting}}\\}
        },
        fixed,
        fixed zerofill={true},
        precision=1,
        multicolumn names=l,
        empty cells with={},
        columns={dataset,setup,browser,sampling-rate, accuracy},
        columns/dataset/.style={column name={}, string type, column type={p{0.18\columnwidth}}},
        columns/setup/.style={column name={}, string type, column type={p{0.16\columnwidth}}},
        columns/browser/.style={column name={}, string type, column type={p{0.16\columnwidth}}},
        columns/sampling-rate/.style={string type, column name={Rate, Hz~}, column type={r}},
        columns/accuracy/.style={column name={\hfill\%}, preproc/expr={100*##1}, column type={r}},
        row predicate/.code={}%
      ]{\dataclassificationbasic}
      \end{threeparttable} 
      \end{table}

In this experiment, we evaluated the classification accuracy of our attack in a so-called \textit{closed-world} scenario,
when the attacker aims to identify the visited website (application) among a predefined list of websites (applications).

For website
fingerprinting, we collected magnetometer and CPU utilization 
traces during retrieval of the \varNumOfWebsites most popular websites from the Alexa
Top 500 Global Sites list~\cite{link-alexa-top500}, merging websites
with multiple domains together
(e.g., google.*). We collected \varNumOfSamples
traces per website, with a duration of 12s each.  Similarly, for
application fingerprinting, we collected traces of \varNumOfApps applications being launched,
\varNumOfSamples traces per application, with a duration of 12s each.
The applications were taken from the list of popular
Android
applications~\cite{link-androidranking}.
A full list of used websites
and applications, as well as classification results for individual
websites and applications, are provided in Appendix~\ref{sec:appendix}. 
All traces were collected on a Google Pixel 2 smartphone lying on a table in the office
environment.

Afterwards, we ran the classification using both sensor and \textit{/proc/stat}
data. The results in terms of classification accuracy are shown in Table~\ref{table:classification-basic}.
As we can see, 
the classifier performs with an accuracy of over 80\% for
website and application fingerprinting. Notably, the proposed
approach has a similar performance in comparison to the
classification based on
actual CPU activity collected through
\textit{/proc/stat}.  These results indicate that the magnetometer-based
side channel leaks sufficient information about CPU activity.

Classification accuracies for different setups are also compared in 
Table~\ref{table:classification-basic}.
More specifically, we separately tested
website retrieval in an embedded WebView component 
with cache disabled, as well as using a full mobile Chrome web
browser with cache enabled. As one can see, the
classification results are similar for both cases. However, the
caching does affect the resulting
patterns.
We also achieved \varWebRecordingAccuracy accuracy 
with web-based recording of sensors using Generic Sensor API,
which proves the applicability of our method to the in-browser scenario.

Additionally, we evaluated the classifier on a larger dataset. We increased the number of websites 
to 100 and repeated the experiment on a single device in the in-app scenario with an embedded WebView component.
The classifier performed with an accuracy of 87.6\%, comparable to 90.5\% in the initial setup
(see Table~\ref{table:classification-basic}). 

\pgfplotstableread[col sep = comma]{tables/classification-devices.csv}\dataclassificationdevices
\begin{table}[t]
\caption{Classification accuracy for website fingerprinting in the in-app scenario for several smartphones, for
intra-device and inter-device modes.}
\label{table:classification-devices}
\centering
\small
\begin{threeparttable}
\pgfplotstabletypeset[
  col sep = comma,
    every head row/.style={before row={%
                              \multicolumn{1}{l}{Device}
                           & \multicolumn{1}{l}{Setup$^\text{a}\!\!$}
                           & \multicolumn{2}{c}{Accuracy, \%} \\
                         },
                         after row=\midrule,
  },
  fixed,
  fixed zerofill={true},
  precision=1,
  multicolumn names=l,
  empty cells with={---},
  columns={device,setup,accuracy,accuracy-crossclass},
  columns/device/.style={column name={}, string type, column type={p{0.38\columnwidth}}},
  columns/setup/.style={column name={}, string type, column type={p{0.1\columnwidth}}},
  columns/accuracy/.style={column name={intra-device}, preproc/expr={100*##1}, column type={c}},
  columns/accuracy-crossclass/.style={column name={inter-device}, preproc/expr={100*##1}, column type={c}},
  row predicate/.code={}%
]{\dataclassificationdevices}
\begin{tablenotes}
\begin{footnotesize}
\item[a] V --- Visual Studio App Center; L --- lab
\end{footnotesize}
\end{tablenotes}
\end{threeparttable}
\end{table}

Afterwards, we ran the website fingerprinting experiment on five other smartphones
in the cloud environment.
We calculated the success rates for
intra-device (with a training and testing
performed on individual devices) and inter-device
(with a training phase performed on traces from all devices, and testing on individual devices)
modes.
The results are
summarized in Table~\ref{table:classification-devices}. 
Google Pixel XL and Samsung smartphones performed worse than
other devices due to the lower sampling rate and the lower SNR ratio,
(see Section~\ref{sec:evaluation:leakage}),
respectively. The activity patterns
are also device-specific. Therefore,
the attacker may need
to train the classifier on numerous devices or take into account the target device model.

\label{sec:evaluation:performance}
\label{sec:evaluation:sampling-rate}

\begin{figure}[t!]
\centering
    \includegraphics[trim={0mm 2mm 0mm 0mm},clip,width=0.65\columnwidth]{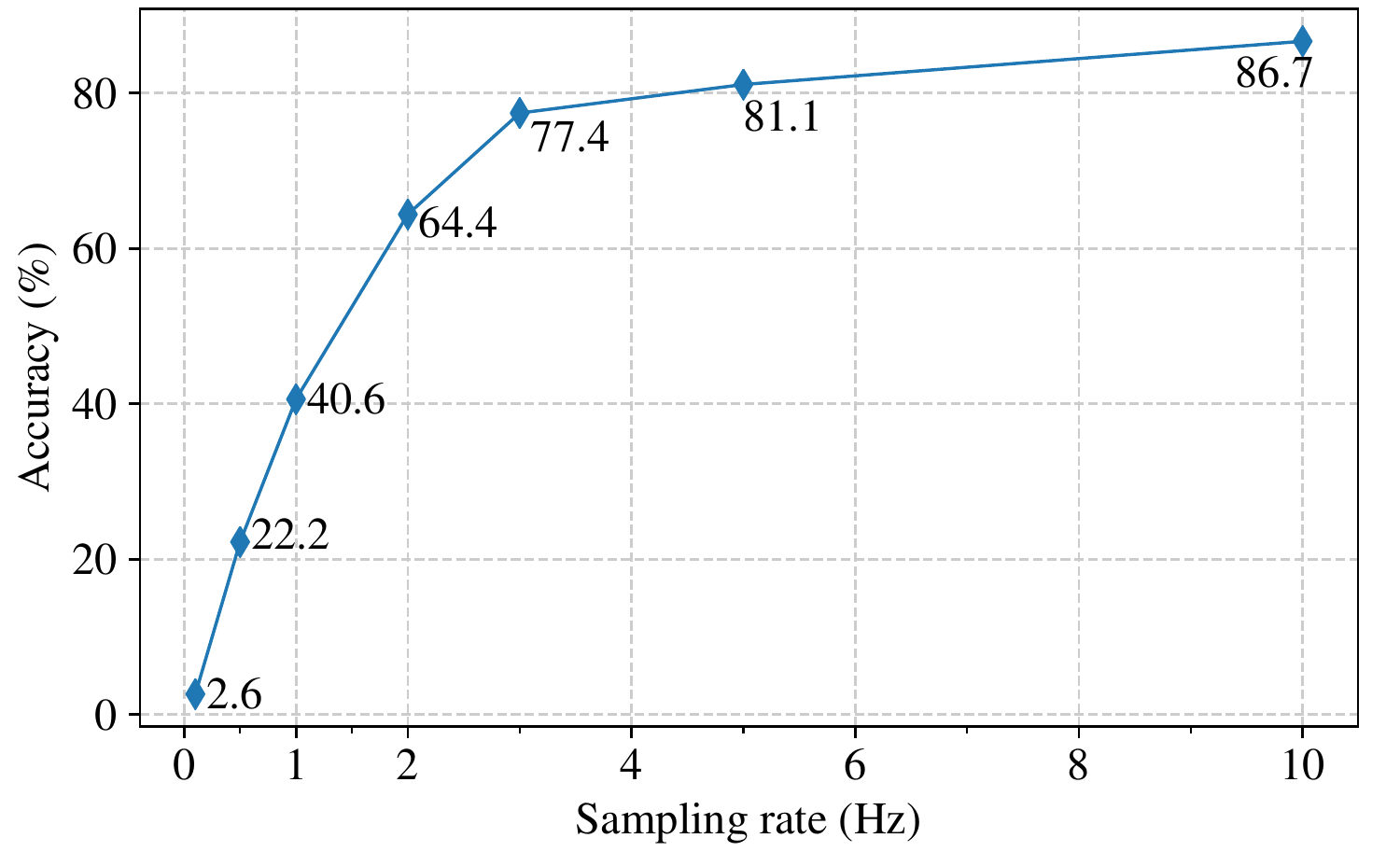}
    \caption{Classification accuracy of website fingerprinting depending on a sampling rate.}
  \label{fig:sampling-rates}
\hfill
\end{figure}

Finally, we
evaluated how the sampling rate of sensor data
gradually affects the classification accuracy.
For this purpose,
we further decreased the sampling rate for the dataset of websites recorded in the in-browser scenario and calculated
the resulting classification accuracies.
Figure~\ref{fig:sampling-rates} shows the results.
As one can see, the sampling rate needs to be reduced to less than 1 Hz in order to make the attack impractical.

\subsection{Open-world scenario}

In this section, we evaluate our classifier
in a so-called \textit{open-world} scenario.
In comparison
to the \textit{closed-world} scenario,
(evaluated in Section~\ref{sec:evaluation:fingerprinting}),
a victim can
visit a much larger set of websites not known to the
attacker. Consequently, the attacker cannot generalize the classifier and identify every visited website.
Instead, the attacker aims to identify whether a victim visits specific websites,
further referred to as \textit{monitored} websites.
To perform the experiment, we first collected traces for \varNumOfWebsites most
popular websites,
\varNumOfSamples traces per website, to train the classifier
similarly to the closed-world scenario.
However, in this case, we selected five the most popular
websites to be five monitored classes.
Other \varOpenWorldNonMonitoredTraining~websites were labeled as \textit{not monitored}, i.e., they belonged to a
separate class. For testing, we used another
list of popular websites~\cite{link-majestic-list}, which is larger than the Alexa list. We
collected one trace for each of the \varOpenWorldNonMonitored~most popular websites,
excluding \varNumOfWebsites websites
(or their alternative domains)
used in the training phase.
Finally, we collected \varOpenWorldMonitored~traces of each of the five monitored websites,
to have a total of \varOpenWorldTotal~traces in the testing set. All traces were
collected on a Google Pixel 2 smartphone in the in-app recording mode.

To evaluate the results, for each class we calculated
the \textit{precision} and \textit{recall} using the following formulae:
$$ precision =  \frac{TP}{TP+FP},~~recall = \frac{TP}{TP+FN},$$
where $TP$ (True Positives) is the number of correctly classified traces of the considered class, $FP$ (False Positives) --- the number of
traces of other classes which are incorrectly classified as the considered class, $FN$ (False Negatives) --- the number of traces incorrectly
classified as the other class.
The previously used
overall classification
accuracy (99.6\% in this case) is not a practical
metric for the open-world experiment, as the classes are
imbalanced, i.e., the number of traces for non-monitored websites
significantly exceeds the number of traces for monitored websites.

Table~\ref{table:classification-openworld} shows the classification
results. We can see that the average achieved recall of \varOpenWorldTP~is lower in comparison to
the closed-world scenario,
but is still practical.
The precision is, however, relatively
high: \varOpenWorldPR~ for all websites and \varOpenWorldPRMonitored~for monitored classes. The high precision
is especially valuable in the open-world scenario, as it ensures
the attacker that the victim did visit the monitored website if it was identified by the classifier.
As a result, we believe that our approach is applicable
to the open-world scenario.

\pgfplotstableread[col sep = comma]{tables/open-world-4.csv}\dataopenworld
\begin{table}[t!]
\caption{Classification results for the open-world scenario, in terms of the precision and the recall
for the five monitored websites.}
\label{table:classification-openworld}
\centering
\small
\begin{threeparttable}
\pgfplotstabletypeset[
  col sep = comma,
  every head row/.style={before row={}, after row=\midrule, },
  every row no 4/.style={before row={}, after row=\midrule, },
  fixed,
  fixed zerofill={true},
  precision=1,
  multicolumn names=l,
  empty cells with={---},
  columns={website,precision,recall,f1-score},
  columns/website/.style={column name={Website}, string type, column type={p{0.42\columnwidth}}},
  columns/precision/.style={column name={Precision,\%}, preproc/expr={100*##1}, column type={p{0.12\columnwidth}}},
  columns/recall/.style={column name={Recall,\%}, precision=1,preproc/expr={100*##1}, column type={p{0.12\columnwidth}}},
  columns/f1-score/.style={column name={F1 score,\%}, precision=1,preproc/expr={100*##1}, column type={p{0.12\columnwidth}}},
  row predicate/.code={}%
]{\dataopenworld}
\end{threeparttable} 
\end{table}

\subsection{Continuous usage}
\label{sec:evaluation:continuous-usage}

In this experiment, we evaluated the ability of the attacker to
detect the starting point of the trace
to be classified
in a
continuous recording stream. The detection is performed
by calculating the cross-correlation
with the predefined pattern,
as we
described in Section~\ref{sec:methodology:continuous-usage}.
We evaluated the approach in the scope of application fingerprinting
and chose the Chrome browser as the target activity.
We made \varCURecordings~continuous recordings lasting \varCUDuration~each,
and within every recording we opened the target application
and two other applications at specific non-overlapping time points.
The applications for each recording were randomly chosen
from the dataset.
This
way, traces contained the pattern
corresponding to the target application, as well as noise from other activities.
For each recording, we calculated the cross-correlation between
the recorded trace and a pattern computed
for the target application.

Then, we detected local
maxima (peaks) in the
result. The set of peaks
was filtered according to three threshold parameters: peak height,
prominence and width.
We considered a peak as true
positive if it was discovered within a \varCUInterval-interval around the time point when the target application
was actually opened.
Other detected peaks were considered as false positives. A false negative was assumed if there was no peak within the corresponding interval.
In the end, we calculated the classification precision and recall
according to these definitions,
to indicate how effectively cross-classification can narrow the search area.
For \varCURecordings~recordings and our set of parameters, we achieved a precision of
\varCUPrecision~and \varCURecall~recall. The attacker can vary parameters of
the cross-correlation to increase the recall at the expense of precision (i.e.,
discover more peaks, including false positives), and vice versa.

\begin{figure}[t!]
\centering
    \includegraphics[trim={0mm 0mm 0mm 0mm},clip,width=0.7\columnwidth]{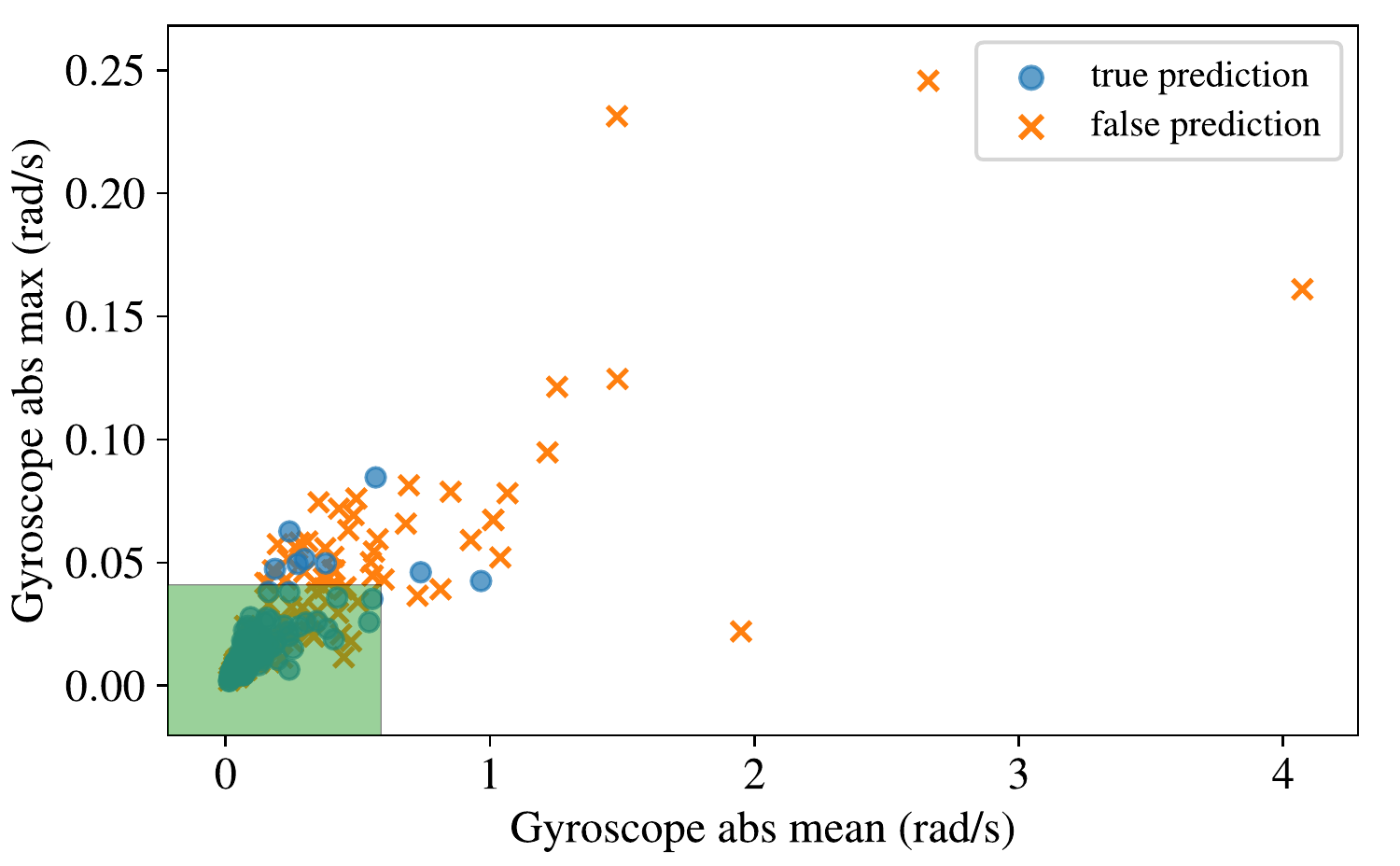}
    \caption{Distribution of traces with regard to their gyroscope-based metrics for movements.
    Numerous wrongly classified traces lie outside the highlighted threshold area.}
  \label{fig:movements}
\end{figure}

Finally, we ran the classification at all discovered
time points including false positives.
This step ensures that the cross-correlation approach in the first step identifies the
time points with sufficient precision, so that the classifier can correctly identify the true positive samples.
The accuracy in our experiment reached \varCUAccuracy, which is
comparable to the \varCUAccuracyCompared 
achieved in the closed-world experiment with a known
beginning point. The decrease is observed due to a number of false positives
at time points corresponding to noise,
as the classifier in the closed-world scenario has not being trained on noise data.
As a result, the experiment shows that the
attacker can efficiently reduce the amount of data to be processed
using the proposed approach.

\begin{table*}[t!]
\caption{Comparison with other related works exploiting side-channel information leakage for website and/or application fingerprinting.
}
\label{table:relatedworks}
\centering
    \footnotesize
\pgfplotstableread[col sep = comma]{tables/related-works-acm.csv}\datarelatedworks
\begin{threeparttable}
\pgfplotstabletypeset[
  col sep = comma,
  every head row/.style={before row={}, after row=\midrule, },
  fixed,
  fixed zerofill={true},
  precision=2,
  multicolumn names=l,
  empty cells with={---},
  columns={work,target,source,platform,defence},
  columns/work/.style={column name={Work}, string type, column type={p{0.2\textwidth}}},
  columns/target/.style={column name={Attack$^\text{a}\!\!$}, string type, column type={p{0.14\textwidth}}},
  columns/source/.style={column name={Leakage source}, string type, column type={p{0.25\textwidth}}},
  columns/platform/.style={column name={Platform}, string type, column type={p{0.17\textwidth}}},
  columns/defence/.style={column name={Blocked}, string type, column type={p{0.10\textwidth}}},
  row predicate/.code={}%
]{\datarelatedworks}
\begin{tablenotes}
\begin{footnotesize}
\item[a] WF --- website fingerprinting; AF --- application fingerprinting
\item[b] Attacks use power traces collected using hardware; prevention is specified for \textit{sysfs} traces 
\item[c] Evaluation is presented for desktop browsers, but potentially generalizes for mobile platforms
\end{footnotesize}
\end{tablenotes}
\end{threeparttable} 
\end{table*}

\subsection{Robustness to movements}
\label{sec:evaluation:movements}

In this experiment, we evaluated the classification accuracy when the smartphone is being held in hand, and
our approach to identify traces affected by movements described
in Section~\ref{sec:methodology:movements}.
We used the classifier trained for the website fingerprinting in the closed-world scenario on a static device (see Section~\ref{sec:evaluation:performance}).
Afterwards, we recorded a total of 500 test traces while freely holding a smartphone in hand.

When the classifier was applied to the whole test dataset without filtering, the overall accuracy dropped to 64.8\%, 
indicating that movements do affect the measurements.
However, wrongly identified traces could be filtered out using the proposed approach: in Figure~\ref{fig:movements},
one can see that for numerous wrongly classified traces the thresholds for indicating movements are
exceeded.
By applying the filtering based on the thresholds before the classification,
21\% of the measurements were identified as affected by movements.
After removing affected traces from the dataset, 
the accuracy reached 73.3\%.
The accuracy is lower in comparison to the accuracy
achieved for the static device (90.5\%), but remains practical. 
Nevertheless, a larger user study and more detailed analysis of the impact of
movements may be needed to prove the wide applicability of the approach.


\section{Related work}
\label{sec:related-work}

\subsection{Website and app fingerprinting on mobile devices}
Researchers
have shown that different side-channel information can be used
to infer applications and websites opened on a smartphone. 
Jana and Shmatikov~\cite{Jana:2012aa} observed the memory footprint of a browser (available through \textit{procfs}) 
to enable website fingerprinting. Zhou~et~al.~\cite{Zhou:2013aa} and Spreitzer~et~al.~\cite{Spreitzer:2016aa} showed that 
the Android data-usage statistics API provides precise information about network activity
and allows to fingerprint applications and websites.
Gulmezoglu~et~al.~\cite{Gulmezoglu:2017aa} used information about system performance counters to establish
website fingerprinting, whereas Diao et al.~\cite{Diao:2016aa} exploited information about system interrupts to establish application fingerprinting, with both leakage sources available through \textit{procfs}.
Several researchers showed that power consumption traces, collected through
\textit{sysfs}~\cite{Chen:2017ab,Yan:2015aa}, using a malicious charger~\cite{Yan:2015aa}
or a malicious battery~\cite{Lifshits:2018aa}, 
are highly correlated with the CPU activity pattern, and therefore, also can be used as leakage source
to infer opened applications~\cite{Chen:2017ab,Yan:2015aa} and websites~\cite{Clark:2013aa,Yang:2017aa,Lifshits:2018aa}.
Recently, Spreitzer~et~al. discovered multiple leakage sources available through
\textit{procfs}~\cite{Spreitzer:2018aa} and Android APIs~\cite{Spreitzer:2018ab}, which allow inferring website
and application activity.
Finally, several works have been presented on microarchitectural side channels, which can be used to infer
information about visited websites~\cite{Oren:2015aa,Lipp:2018aa}. In the most recent work,
Shusterman~et~al.~\cite{Shusterman:2019aa} demonstrated the cache
occupancy side channel to establish website fingerprinting in the in-browser scenario.
Although the results were
presented for the desktop platform, the approach may be applied to mobile devices.

Table \ref{table:relatedworks} summarizes these prior works and compares them with our approach.
As we can see, most of the leakage sources are already blocked in the latest Android OS.
Furthermore, currently available \textit{procfs} resources can be blocked in future versions of Android without serious impact
on existing applications as they provide system-specific technical information.
In contrast, our attack works on the latest Android 9 and access to magnetometer cannot be completely blocked, since numerous
applications rely on magnetometer values (e.g., navigation applications).
Furthermore, almost all prior works require a malicious application to be installed on a device, while our attack can be launched from a web page.

\subsection{Exploiting the reaction of magnetometers to EM activity}
The reaction of magnetometers to electromagnetic activity emitted by computer components
has been used to establish inter-device covert channels. Researchers used magnetometers to receive covert signals
from a nearby computer encoded into hard drive activity~\cite{Biedermann:2015aa}, CPU activity~\cite{Guri:2018ab}, and combined
I/O activity~\cite{Matyunin:2016ab}.
Matyunin~et~al.~\cite{Matyunin:2018ab} proposed a magnetometer-based \textit{intra-device} covert
channel
on smartphones. The authors demonstrated that the magnetometer can be affected by the peak CPU activity,
emitted by a webpage. 
In a recent work~\cite{Cheng:2019aa}, Cheng~et~al. exploited the reaction of magnetometers to EM activity 
to infer applications and webpages opened on victim's laptop located in victinity to the attacker's smartphone.
In this work, we show that magnetometer disturbance on mobile devices accurately represents the patterns of the \textit{internal} CPU activity,
evaluate this effect on a large number of modern devices, and show that a malevolent application on a victim's smartphone
can infer running activity, namely, to perform application and website fingerprinting.


\section{Countermeasures}
\label{sec:countermeasures}

There are several possibilities to prevent the presented information leakage through
magnetometer disturbance: 
\begin{itemize}
\item Physical shielding with
ferromagnetic materials is the most straightforward way to limit
the susceptibility of the sensor to electromagnetic activity. However,
this measure opposes an industry trend of making smartphones
thinner and lighter, and cannot protect existing devices from the
attack.  
\item As we have discovered in our
experiments in Section~\ref{sec:evaluation:leakage}, some smartphones and
tablets actually do not react to CPU activity, presumably due to the
sensor location relative to the CPU or power supply components. We,
therefore, believe that the location of the sensors should be 
taken into account when designing the layout of the smartphone motherboard.
\item Based on our evaluation in Section~\ref{sec:evaluation:sampling-rate},
further limiting the sensor sampling rate to 1~Hz significantly
reduces the classification accuracy of fingerprinting. However, with such a lower sampling rate
it may be still possible to infer information about
more coarse-grained activities. Furthermore, it may negatively affect the performance of legitimate
applications.
\item An explicit user permission can be introduced to limit access to magnetometers.
However, users may not correctly perceive potential privacy threats emerging from sensors in
mobile devices~\cite{Mehrnezhad:2018aa}. Therefore,
an explanation of potential risks might be needed. Moreover, a lot
of mobile devices in use run outdated operating system
versions~\cite{link-android-outdated}.
\item To limit the attack surface of our attack, access to
magnetometers can be restricted for applications opened in the split-screen mode
and can immediately be blocked when the application goes to the background.
\end{itemize}

The described countermeasures would require hardware or software
changes, may have performance or production cost drawbacks, and
require careful design decisions. 
In particular, we are
concerned about the ongoing deployment of the Generic Sensor API
in browsers as access to the magnetometer from the browser significantly
extends the attack surface.
We recommend to require an explicit
permission to access the magnetometer on web pages.
An alternative recommendation would be to
further reduce the sampling rate.


\section{Discussion}
\label{sec:discussion}

In this section, we discuss some related aspects and
directions for future work.

First, as shown in other works on website fingerprinting
(e.g., ~\cite{Juarez:2014aa,Wang:2013aa}),
aging of sampling data affects classification accuracy.
Therefore, the attacker needs to repeat the learning phase
periodically. One interesting direction for future work would be a detailed investigation
of which elements on a web page affect the classification the most when
being changed, for different fingerprinting methods. For example, increasing the web page
size by extending the text content can affect fingerprinting based
on traffic analysis, but may have no substantial
effect on the sensor disturbance in our approach, since text rendering
is computationally inexpensive for the CPU. 


Second, in principle, magnetometer sensors are suscpetible to external
electromagnetic noise. However, as shown in other works~\cite{Guri:2018ab,Matyunin:2016ab},
magnetometers are affected by the noise from nearby computers only at
short distances ($\le$15cm). We performed all experiments in a typical office environment with
natural arrangement of multiple electronic devices, such as laptops, WiFi access points, and other smartphones.
As our results indicate, activity of these devices did not impair the performance of our approach.
Nevertheless, systematic analysis of potential external noise sources and their impact on classification 
can be performed as future work.



Finally, it would be interesting to combine our approach with works
exploiting other side-channel information on smartphones,
especially leakages of other nature such as memory access statistics.
In this way, a feature set combining different side-channel information
can potentially improve the classification accuracy.

\section{Conclusion}
\label{sec:conclusion}

In this work, we presented a method to identify running applications and websites 
on mobile devices based on the reaction of magnetometers
to the internal CPU activity. We observed
this side channel on a large number of modern devices and demonstrated
that this information leakage is sufficient to identify opened
websites and applications. The presented method
does not require any user permissions and can be run
in both in-app and in-browser scenarios,
posing a significant threat against the privacy of mobile users.

    \bibliographystyle{ACM-Reference-Format}
    \bibliography{IEEEabrv,bibliography/sc-magnetic} 
    \balance


\newpage
\onecolumn

\appendix
\section{Appendix}
\label{sec:appendix}

 \begin{table*}[h!]
        \caption{Full list of mobile devices on which sensor measurements 
        correlate with CPU activity.
        }
        \label{table:snr-affected-full}
              \centering
    \footnotesize
        \pgfplotstableread[col sep = comma]{./tables/snr-3-affected.csv}{\datasnraffected}
        \begin{threeparttable}
          \renewcommand{\arraystretch}{0.70}
        \pgfplotstabletypeset[
          col sep = comma,
            every head row/.style={before row={%
                                      \multicolumn{1}{c}{Smartphone}
                                   & \multicolumn{1}{c}{Setup$^\text{a}\!\!$}
                                   & \multicolumn{1}{c}{Magnetometer$^\text{c}\!\!$}
                                   & \multicolumn{2}{c}{Correlation}
                                   & \multicolumn{1}{c}{SNR,} \\
                                 },
                                 after row=\midrule,
          },
          every row no 0/.style={
            before row={\multicolumn{6}{l}{\textbf{Android}}\\},
          },
          every row no 38/.style={
            before row={\midrule\multicolumn{6}{l}{\textbf{iOS}}\\},
          },
          fixed,
          fixed zerofill={true},
          precision=2,
          multicolumn names=l,
          empty cells with={---},
          columns={smartphone, setup, magnetometer, correlation, correlation-cpu, snr},
          columns/smartphone/.style={column name={}, string type, column type={p{0.30\columnwidth}}},
          columns/setup/.style={column name={}, string type, column type={p{0.10\columnwidth}}},
          columns/magnetometer/.style={column name={}, string type, column type={l}},
          columns/correlation/.style={column name={~Pattern~}, column type={r}},
          columns/correlation-cpu/.style={column name={~CPU$^\text{b}\!\!$~}, column type={r}},
          columns/snr/.style={column name={dB}, precision=1,column type={r}},
          row predicate/.code={}%
        ]{\datasnraffected}
        \begin{tablenotes}
        \begin{footnotesize}
        \item[a] V --- Visual Studio App Center; A --- AWS Device Farm; L --- lab
        \item[b] CPU utilization data is available only on devices running Android $\le$ 7 (over \textit{/proc/stat}) and iOS (over \textit{host\_processor\_info}).
        \item[c] For Android devices, information provided by the \textit{Sensor} API. For iOS devices, information from publicly available online resources is used.
        \item[] 
        \end{footnotesize}
        \end{tablenotes}
        \end{threeparttable} 
        \end{table*}

        \newpage
        \begin{multicols}{2}
\pgfplotstableread[col sep = comma]{tables/websites.csv}\datawebsites
\begin{table}[H]
\caption{Classification results for websites.}
\label{table:websites}

\centering

    \footnotesize
\begin{threeparttable}

          \renewcommand{\arraystretch}{0.97}
\pgfplotstabletypeset[
  col sep = comma,
  every head row/.style={before row={}, after row=\midrule, },
  every last row/.style={before row=\bottomrule},
  fixed,
  fixed zerofill={true},
  precision=2,
  multicolumn names,
  empty cells with={---},
  columns={website, precision, recall, f1-score},
          columns/website/.style={column name={Website}, string type, column type={p{0.15\textwidth}}},
  columns/precision/.style={column name={Precision}, column type={r}},
  columns/recall/.style={column name={Recall}, column type={r}},
  columns/f1-score/.style={column name={F1 score}, column type={r}},
  row predicate/.code={}%
]{\datawebsites}
\end{threeparttable} 
\end{table}

\vspace{\fill}
\columnbreak

\pgfplotstableread[col sep = comma]{tables/apps.csv}\dataapps
\begin{table}[H]

\caption{Classification results for applications.}
\label{table:applications}

\centering
    \footnotesize
\begin{threeparttable}
          \renewcommand{\arraystretch}{0.75}
\pgfplotstabletypeset[
  col sep = comma,
  every head row/.style={before row={}, after row=\midrule, },
  every last row/.style={before row=\bottomrule},
  fixed,
  fixed zerofill={true},
  precision=2,
  multicolumn names,
  empty cells with={---},
  columns={application, precision, recall, f1-score},
  columns/application/.style={column name={Application}, string type, column type={p{0.27\textwidth}}},
  columns/precision/.style={column name={Precision}, column type={r}},
  columns/recall/.style={column name={Recall}, column type={r}},
  columns/f1-score/.style={column name={F1 score}, column type={r}},
  row predicate/.code={}%
]{\dataapps}
\end{threeparttable} 
\end{table}
\end{multicols}

 \begin{table}[t!]
        \caption{List of mobile devices and their magnetometers not affected by their CPU activity.}
        \label{table:snr-not-affected}
        \centering
    \footnotesize
        \pgfplotstableread[col sep = comma]{tables/snr-3-not-affected.csv}{\datasnrnotaffected}
        \begin{threeparttable}
        \renewcommand{\arraystretch}{0.95}
        \pgfplotstabletypeset[
          col sep = comma,
          every head row/.style={before row={}, after row=\midrule, },
          every row no 0/.style={
            before row={\multicolumn{3}{l}{\textbf{Android}}\\},
          },
          every row no 21/.style={
            before row={\midrule\multicolumn{3}{l}{\textbf{iOS}}\\},
          },
          fixed,
          fixed zerofill={true},
          precision=2,
          multicolumn names=l,
          empty cells with={---},
          columns={smartphone, setup, magnetometer},
          columns/smartphone/.style={column name={Smartphone}, string type, column type={p{0.40\columnwidth}}},
          columns/setup/.style={column name={Setup}, string type, column type={l}},
          columns/magnetometer/.style={column name={Magnetometer}, string type, column type={l}},
          row predicate/.code={}%
        ]{\datasnrnotaffected}
        \end{threeparttable} 
        \end{table}


\end{document}